# Influence of Functionalized Fullerene Structure on Polymer Photovoltaic Degradation

Brian H. Johnson[a], Enaanake Allagoa[a], Robert L. Thomas[a,b], Gregory Stettler[b], Marianne Wallis[b], Justin H. Peel[a], Thorsteinn Adalsteinsson[b], Brian J. McNelis[b], Richard P. Barber, Jr.[a,*]

[a]Department of Physics, Santa Clara University, Santa Clara, CA 95053

[b]Department of Chemistry and Biochemistry, Santa Clara University, Santa Clara, CA 95053

**Abstract**

The time dependence of device performance has been measured for photocells using blends containing the conjugated polymer, poly[2-methoxy-5-(2-ethylhexyloxy)-1,4-phenylenevinylene] (MEH-PPV) with two different functionalized $C_{60}$ electron acceptor molecules: commercially available [6,6]-phenyl $C_{61}$ butyric acid methyl ester (PCBM) or [6,6]-phenyl $C_{61}$ butyric acid octadecyl ester (PCBOD) produced in this laboratory. Performance was characterized by the short-circuit current output of the devices, with the time dependence of PCBM samples typically degrading exponentially. Variations in the characteristic lifetime of the devices were observed to depend on the molar fraction of the electron acceptor molecules (calculated with respect to the MEH-PPV monomer fraction). In comparison to the PCBM samples, the stability of the PCBOD blends was significantly enhanced, with a one or two order of magnitude improvement. Corresponding spectroscopic data with similar time evolution as the transport measurements suggests an independent means for determining and understanding degradation mechanisms.

*Corresponding author. Tel.: +1 408-554-4315; fax +1 408-554-6965
E-mail address: rbarber@scu.edu (R. P. Barber, Jr.)

# 1. Introduction

After almost two decades, polymer photovoltaics continue to be widely investigated given their potential to provide an inexpensive and mechanically robust alternative to Si-based technologies [1,2]. The two major obstacles in the implementation of these materials into commercially-available solar cells are relatively poor power conversion efficiencies and device lifetime. Much of the previous research has focused on the device fabrication and synthesis of polymers to improve efficiencies, which have been optimized at 6% or higher [2]. While lower device efficiency could be considered a tradeoff with decreased cost of the polymer-based devices, lifetime is a critical question in determining the viability of organic photovoltaics. Recent progress has been made in demonstrating manufacturing techniques [3] and packaging to maintain stability during inter-laboratory testing [4]. However in order to advance the technology to the commercial product stage, the stability of these devices must continue to improve.

Work by Krebs and co-workers has highlighted the need for systematic investigation of stability [5-7]. A standardized set of observations and measurements are suggested as a starting point for this work. Previous investigations have probed the effects of morphology [8-10], annealing [8,10], architecture [5], and electron acceptor concentration on device lifetime [11-13]. In particular, it has been found that $C_{60}$ concentration in the MEH-PPV blend improved the lifetime by an order of magnitude [12]. Other work has studied the variations in the functionalized $C_{60}$ electron acceptor molecules and the effects on device performance [14,15]. In this study our goal is to elucidate the role of $C_{60}$ concentration in device stabilization and to examine modifications to the $C_{60}$ additive which have lead to measured differences in the degradation process. Data showing these differences will be presented and compared.

## 2. Experimental

Devices were prepared using standard solution processing techniques inside an inert-atmosphere glove box. Indium-tin-oxide (ITO) pre-coated glass substrates were patterned and then spin-coated with poly[3,4-ethylenedioxythiophene]:poly[styrenesulfonate] (PEDOT:PSS) and baked at 200 ºC. The active layer, a blend of MEH-PPV with either PCBM or PCBOD was then spun from a chlorobenzene solution onto the sample. The weight fractions of the solutions were between 0.6 and 1.0 percent, with spin speeds of roughly 1000 rpm. Previous work suggests that the resultant film thicknesses are likely of 100-150 nm [16], but direct measurements are not available for the present samples. The two functionalized $C_{60}$ molecules are shown in Fig. 1. PCBM is the electron acceptor widely used in polymer photovoltaic measurements and was acquired commercially. PCBOD was synthesized in this laboratory from [6,6]-phenyl $C_{61}$ butyric acid coupling with octadecanol using previously reported methods [17,18]. The chemical structure was corroborated using proton nuclear magnetic resonance spectroscopy and fast atom bombardment mass spectroscopy. The device preparation is concluded by evaporating ~1 nm of LiF followed by 100 nm of Al to form the top electrodes in a standard bell-jar evaporator system equipped with a quartz crystal thickness monitor. Since this apparatus is not integrated into the glove box, samples were transferred using a vacuum tight vessel containing glove box atmosphere. Insertions into the evaporator require brief exposure to atmospheric conditions. The elapsed time between the initial ambient air exposure until the evaporator pressure of $10^{-1}$ Torr was reached was typically 5 minutes. A schematic of the sample structure is shown in Fig. 2.

Physical measurements were done in ambient atmosphere immediately after removing the sample from the evaporator. With our device layout, each sample typically produced 8 individual photovoltaic devices. Current-voltage (*I-V*) characteristics of the devices were measured alternately in darkness and illuminated by a PV Measurements, Inc. Small-Area Class-B Solar Simulator. At the beginning of each sequence, each of the devices on a sample was briefly tested. The device with the best performance was then measured repeatedly for hours or days depending on its functioning lifetime. The devices near the center of the sample usually performed comparably (current variations within about 30%), with devices near the edges showing higher variability in performance and occasional failure. Edge damage due to mounting in the evaporator and general handling as well as spin uniformity are likely contributors to this effect. Illumination was used only during the actual *I-V* measurements; otherwise the device remained in ambient but low light conditions (no measurable short-circuit current could be detected). The laboratory temperature was controlled at 23-26 ºC with a relative humidity range of 50-70% (not directly controlled).

## 3. Results and Discussion

Typical *I-V* characteristics for a PCBM/MEH-PPV blend device are plotted in Fig 2. For this particular sample the molar fraction, *x*, used was 0.59 where

$$x = \frac{m_{C60} / MW_{C60}}{m_{C60} / MW_{C60} + m_{MEH-PPV} / MW_{MEH-PPV}}$$

with *MW* denoting the molecular weight of each species and *m* denoting its mass. The weight average molar mass of the polymer is given by the manufacturer to be 40,000-70,000 g/mol.

Used here instead is the molar mass of the MEH-PPV monomer. In other words, $x=0.5$ represents a blend with one $C_{60}$ molecule per MEH-PPV monomer.

Also in Fig. 2 it is noteworthy that $V_{OC}$ remains relatively unchanged during the degradation of the current, suggesting that the chemical potential across the device is not changing significantly. $I_{SC}$, presumably a measure of how many active channels exist in the device, is then chosen as the figure of merit to quantify the degradation. Fig. 3 displays the $I_{SC}$ results from the Fig. 2 data as a function of time. In addition, an exponential fit is shown in Fig. 3, supporting a simple qualitative description of an exponential degradation with a characteristic time, $\tau$, in agreement with previous work [12]. The first point in the figure is not included in the fit, since this point was taken during the testing of all devices on the sample. The present configuration of the sample mount does not allow for reproducibly positioning the sample in the solar simulator once it has been moved. All subsequent *I-V* curves were taken with the sample position fixed.

Light absorption using UV-visible spectroscopy was used in parallel to electron transport measurements. Because of the opacity of the Al electrodes, devices were fabricated without the final LiF/Al layer specifically for this purpose. Fig. 4 shows results from an $x=0.8$ sample of PCBM. Plotted is the differential absorption measurement for the *325-600 nm* range for increasing time as indicated by the arrow. The initial absorption scan through the device was used as the base line for the spectra. Decrease in absorption will therefore register as a negative value, whereas increase in absorption will register as a positive value. The dominant feature in these spectra compares very closely with spectra taken of PCBM [19], although the reason for this result is not yet apparent. The 352 nm peak is observed to grow and saturate after about 10-12 hours. The difference spectra also show a slight red shift in the absorption peak, which could indicate increase in scattering within the film or a shift in the dielectric environment of the polymer film due to aging. Since the number density of PCBM cannot grow during the experiment, the increased absorption at 352 nm is likely indicative of increase in efficiency for the absorption, which may point to change in the dielectric environment. By plotting the difference between the saturated spectrum and the time-evolving spectra (inset of Fig. 4), exponential time dependence is revealed. In other words, the saturation time dependence can be modeled as $(1-\exp(-t/\tau))$. From the plot the time constant, $\tau$ is easily determined.

Systematic measurements of the degradation of samples of varying PCBM molar fraction were characterized by the decay of $I_{SC}$ as shown in Fig. 5. These data have been normalized to the *t=0* value for clarity. Previous studies have indicated the enhanced stability of blends as PCBM was added [11-13]. The exponential behavior of these data (appears linear on this semi-log plot) is consistent with the result from this earlier work [12]. Fig. 6 summarizes the results of both the degradation and the initial performance as a function of stoichiometry. For two sets of samples (open triangles and circles), the upper frame shows the initial $I_{SC}$ before significant degradation occurs. The bottom frame shows the corresponding time constants $\tau$ for these devices. The best performing devices are at about *x=0.6* in good agreement with previous results [16], while the best stability occurs with the highest PCBM content. In addition to transport measurement time constants, the UV-vis saturation time constant (star) as derived from data of Fig. 4 is also included. This good agreement suggests that corresponding measurements of the degradation can be made using both transport and UV-vis spectroscopy.

The exponential decay as observed in both the transport and spectroscopic measurements suggests a simple model of the degradation process. Assume there is some number of photovoltaic channels (*N*) contributing to the current output of a particular device, with each of these channels having identical characteristic lifetime. It should be expected that the rate at which they fail should be proportional to their number and the increment of time *dt*. In other words *dN=-Ndt*. This relationship will lead to the obvious exponential dependence. Morphological models for this family of solar cells with many parallel paths for photovoltaic activity would seem to be consistent with this picture [20-22]. In addition, other work has modeled the time dependence with double exponentials [5].

The next phase of these experiments was to determine how changing the electron acceptor molecule would affect both device stability and performance. The choice of PCBOD was made based on a number of factors. Unlike the conducting polymer, structural changes to the fullerene additive are relatively straightforward and therefore have been the focus of numerous synthetic investigations [14-18]. In addition, we rationalized that both device performance and degradation are directly related to aggregation phenomenon during fabrication *and* the lifetime of the device. Given the rapid degradation of the PCBM devices, it seemed to us that these devices must be undergoing a reorganization of the components. Although oxidative breakdown of one or more of the components could be a contributing factor, the rate of degradation clearly suggests a more fundamental change in the device structure. Given the known self-aggregation of fullerenes, we chose a long hydrocarbon chain; such a structure might assert its own self-organization preferences in device fabrication. In addition, the device stability would also be affected since this structural change might decrease the mobility of the functionalized $C_{60}$. Interestingly, numerous fullerenes additives have been prepared containing long-hydrocarbon chains for solubility purposes, only one study has examined the affect of the chain on device performance and that study did not examine device lifetimes [15].

An example of the degradation of a device using the PCBOD electron acceptor molecule is presented in Fig. 7. In contrast to the exponential decay of the PCBM-based devices, a dramatically different time dependence emerges. Note that these data are also plotted as a semi-log, however using the logarithm of *time* and not the current to produce a linear plot. In other words the decay in this sample is logarithmic in time and not exponential as in the PCBM samples. None of the PCBOD samples followed a simple exponential trend as did almost all of the PCBM devices. Instead there is an initially rapid degradation followed by a very slow nearly saturating behavior. Also, measurable currents for the PCBM devices were usually completely gone within a day, whereas this particular PCBOD sample continues to function (albeit poorly) for roughly one month.

UV-vis absorption spectra were also taken for a PCBOD sample from the same blend solution as the sample in Fig. 7. Fig. 8 shows a dramatically different spectral fingerprint of decreasing absorption. In contrast to the absorption spectra seen in Fig. 4 corresponding to PCBM spectra, these results look more related to MEH-PPV spectra as displayed in other work [19]. As with the saturating $C_{60}$ feature in the PCBM samples, this effect is still under investigation. The isosbestic points in the spectrum suggest a direct shift of one type of chemical environment to another. As before the arrow of time is denoted in the figure. Choosing the peak at about 500 nm, the time dependence is plotted in Fig. 9 for about one day duration for this sample. A good correlation with the transport measurements from Fig. 7 is apparent, suggesting that the observation of the decay in photovoltaic function can be revealed using both transport and optical means.

In the PCBOD samples, temporal agreement is again seen between transport and spectroscopic measurements. This correlation suggests that both measurements are a reliable and consistent way to characterize device degradation. The logarithmic time dependence in the PCBOD sample is in dramatic contrast to the exponential result for PCBM. In particular *log(t)* is suggestive of a diffusive process in a 2-dimensional system (see [23] for an example). With longer alkyl chains, the self-diffusion rate of the molecule should be reduced in the film, and the tendency for the molecules to aggregate or de-aggregate may also be reduced. This situation would lead to a device degradation behavior that is more indicative of diffusion, hence the *log(t)* behavior. Optical studies have revealed the importance of aggregation in device performance [22, 24], a result which would be consistent with this picture. A simple planar geometry photovoltaic device showing similar time dependence has been measured by Krebs *et al.* [5]. In their type 4 device, the $C_{60}$ molecules were sequestered in a distinct layer. Such geometry might be expected to produce a diffusive behavior upon time evolution. Although they modeled all of their data in this work consistently with double exponential fits, the time dependence of the current shown in their Fig. 3 is well-described as *log(t)*.

## 4. Conclusions

This work has focused on the degradation in device performance of both PCBM- and PCBOD-based polymer photovoltaic devices. The results are in agreement with the stabilization effects of the addition of the functionalized $C_{60}$ molecule. A dramatic change in the time dependence has been observed in the degradation using an octadecyl ester of the functionalized $C_{60}$ rather than the methyl ester, PCBM. Specifically *log(t)* was observed in the PCBOD as opposed to the *exp(-t)* dependence for PCBM, typical of these types of devices in the literature. Furthermore, the results have shown that both standard transport measurements and spectroscopic measurements appear to reveal the same time dependence of the degradation of devices. Further work will explore this connection and to utilize it to find more promising polymer-fullerene blends, focusing on additional fullerene modifications that will impart improved device stability.

## 5. Acknowledgements

This project has been supported by Santa Clara University funding including a Faculty-Student Research Program Grant, a Center for Nanostructures Grant, a Faculty Development Internal Grant, and a Technology Steering Committee Grant. We acknowledge valuable discussions with D. Romero and the invaluable technical support of S. Tharaud.

**Figure Captions**

Fig. 1. Electron acceptor molecules used in this study: a) [6,6]-phenyl $C_{61}$ butyric acid methyl ester (PCBM) and b) [6,6]-phenyl $C_{61}$ butyric acid octadecyl ester (PCBOD).

Fig. 2. Typical current-voltage (*I-V*) curves showing the degradation of a device in ambient conditions. The arrow indicates the progression of time and the ordinate values at *V=0* are short circuit current $I_{SC}$. *Inset:* a schematic of the device architecture used in this study.

Fig. 3. $I_{SC}$ as a function of time from the *I-V* data in Fig. 2. The fitted curve is a single exponential that yields a time constant used to characterize the degradation lifetime of this device.

Fig. 4. Differential UV-vis absorption spectrum for an x = 0.8 PCBM device evolving with time (denoted by the arrow). The *352 nm* absorption is seen to saturate after long times (about 10 hours). Inset: saturated absorption value (abs($\infty$)) minus the time dependent absorption (abs(t)). The roughly linear dependence of this semi-log plot suggests a $1-\exp(t/\tau)$ behavior. The associated time constant $\tau$ is 186 minutes.

Fig. 5. Semi-log plot of $I_{SC}$ as a function of time for a series of PCBM devices. These curves are normalized to the value of $I_{SC}$ at *t=0*. The dotted lines represent the fit for these data and yield the decay time constants for each device. The enhancement of the stability of the devices as the PCBM molar fraction is increased is apparent.

Fig. 6.  Data for two series of PCBM devices (open symbols; circles are from Fig 5.)  We plot a) the initial $I_{SC}$ and b) the decay time constant as a function of PCBM molar fraction.  The star indicates the time constant derived from the saturation of the 352 nm absorption peak in Fig. 4.

Fig. 7.  The time dependence of $I_{SC}$ for a PCBOD-based device.  The time axis is plotted as log(t), representing a vastly different time behavior from the exp(-t) of PCBM-based cells.

Fig 8.  Differential UV-vis absorption spectrum for a 0.6 MF PCBOD device evolving with time (denoted by the arrow).  Noteworthy is the completely different spectral signature compared to the PCBM data (Fig. 4.)

Fig. 9. The *500 nm* feature from Fig. 8 plotted for roughly one day using the right ordinate axis for the scale. One day from the $I_{SC}$ data of Fig. 7. with the right ordinate axis for its scale. This comparison was made by simply matching the *t=0* values from both data sets and simply scaling overall to roughly match them at *1300 minutes*.

a)

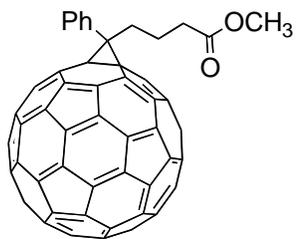

**PCBM**

b)

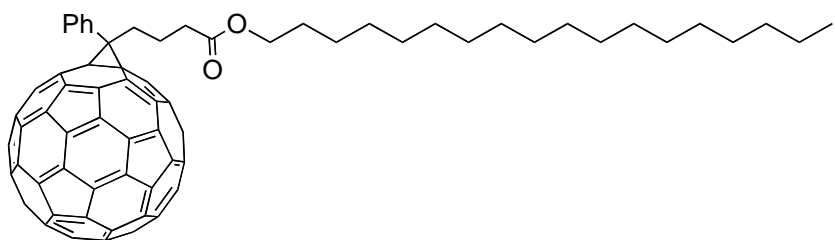

**PCBOD**

Fig 1

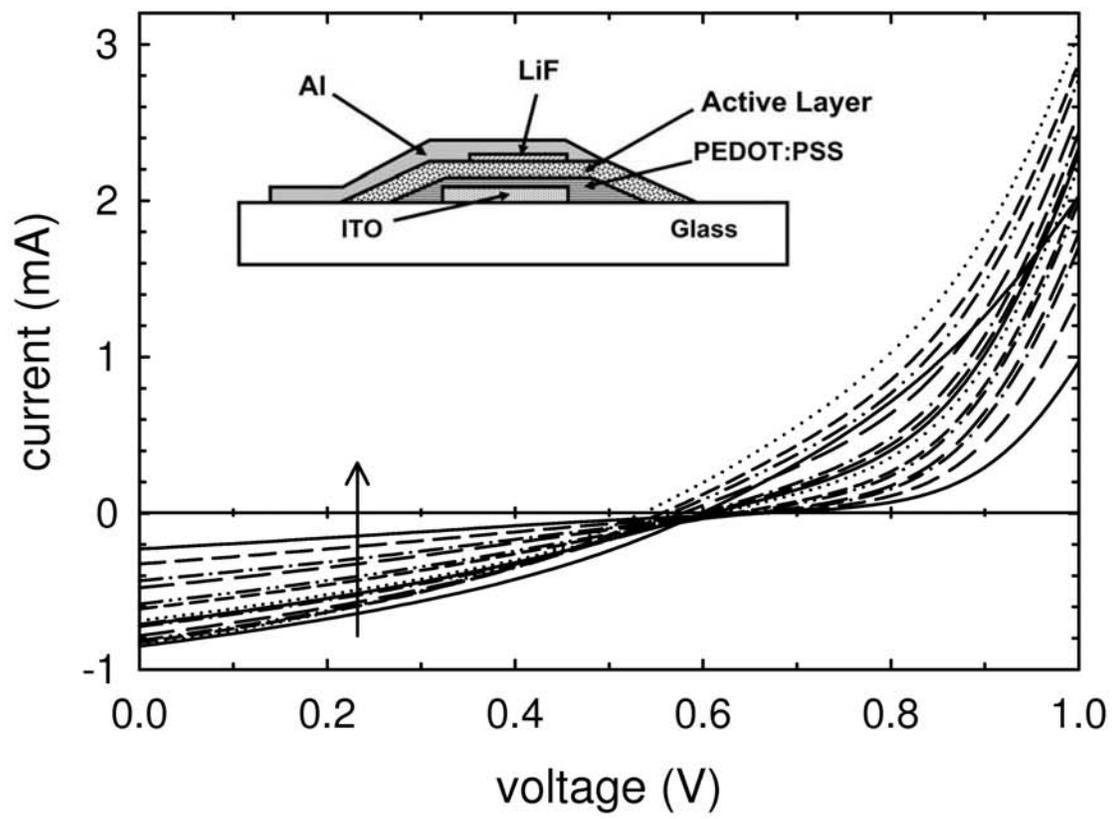

Fig 2

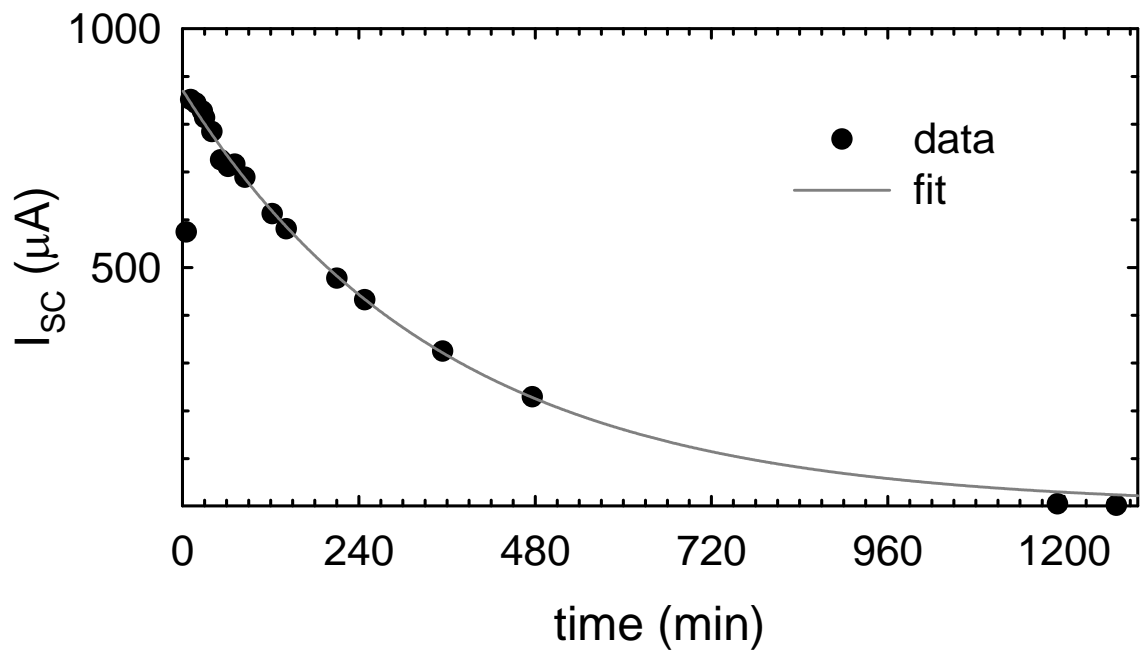

Fig 3

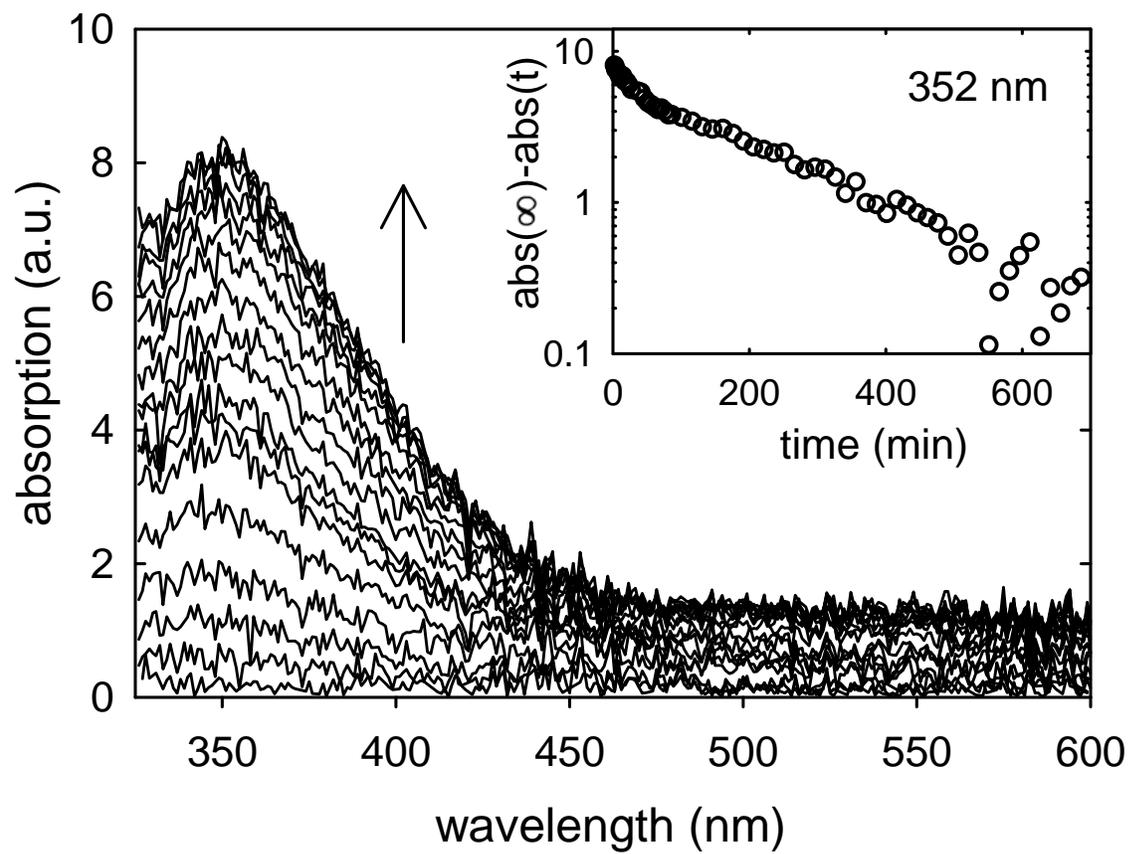

Fig 4

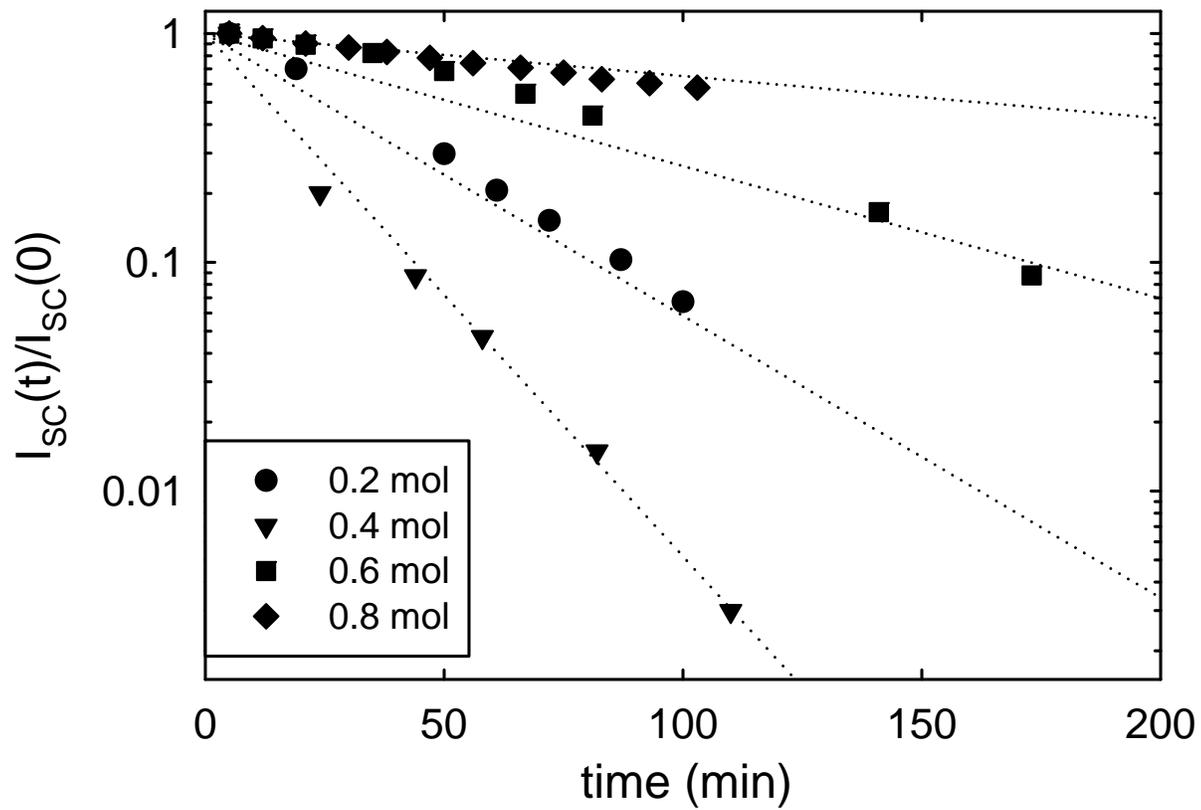

Fig 5

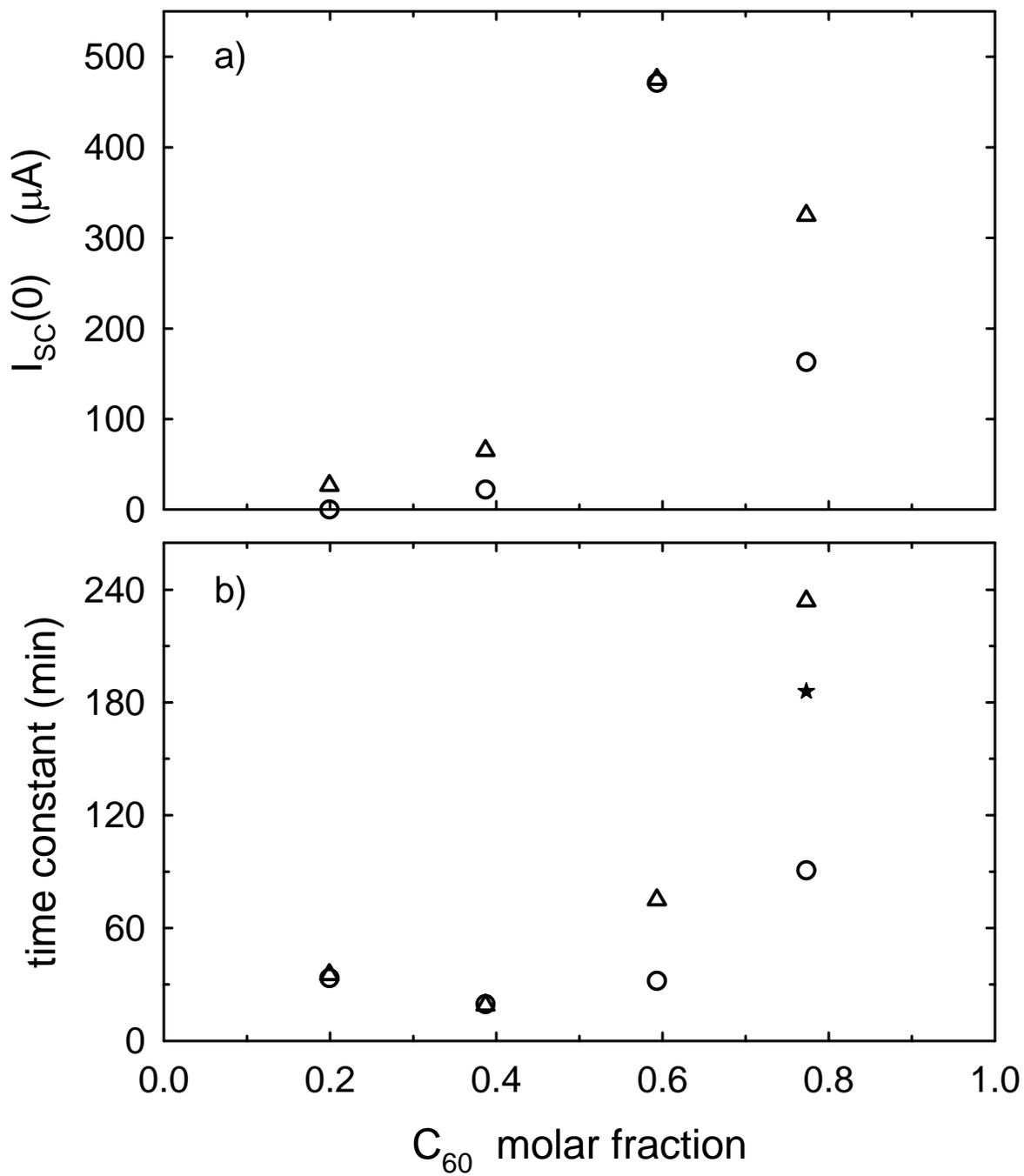

Fig 6

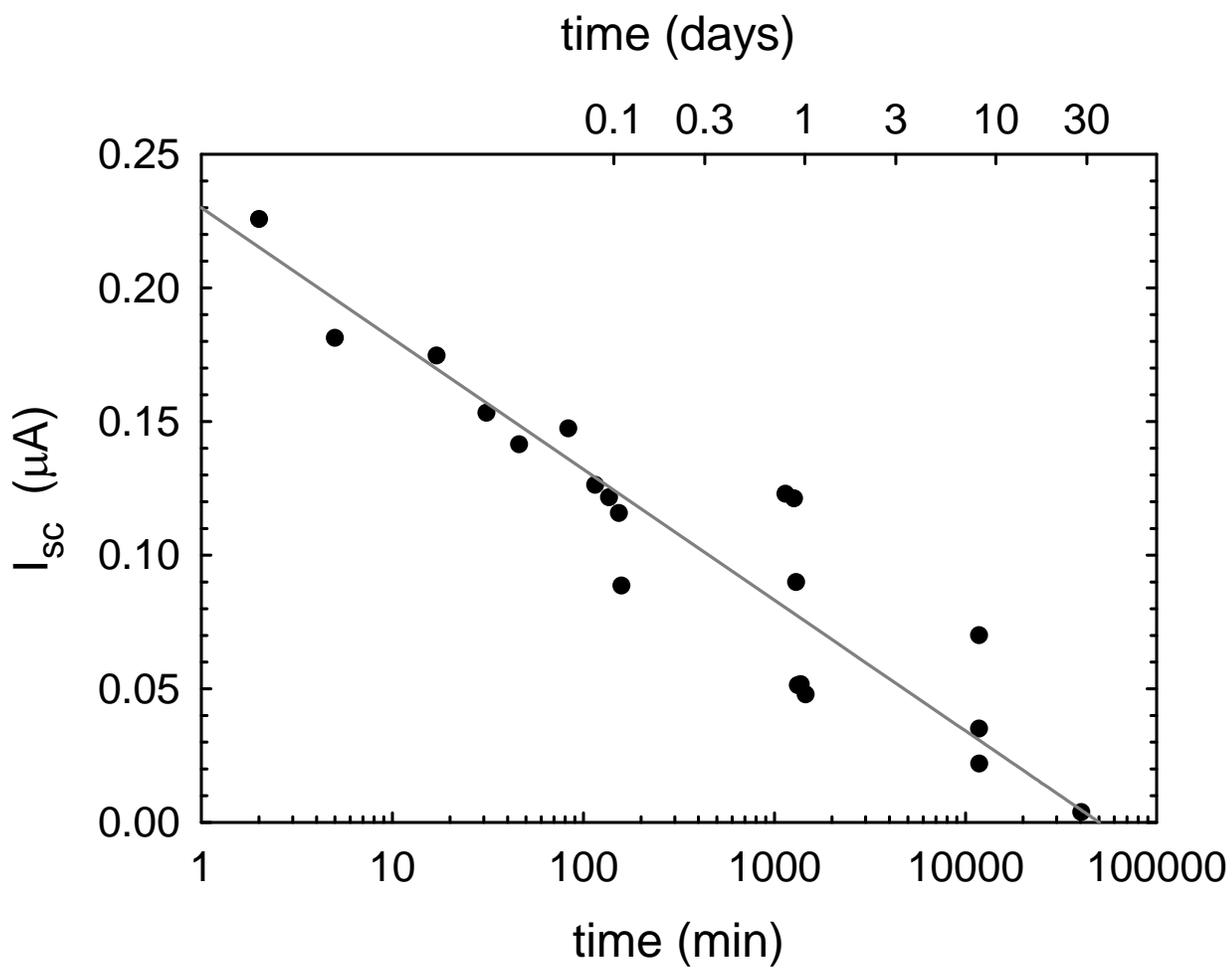

Fig 7

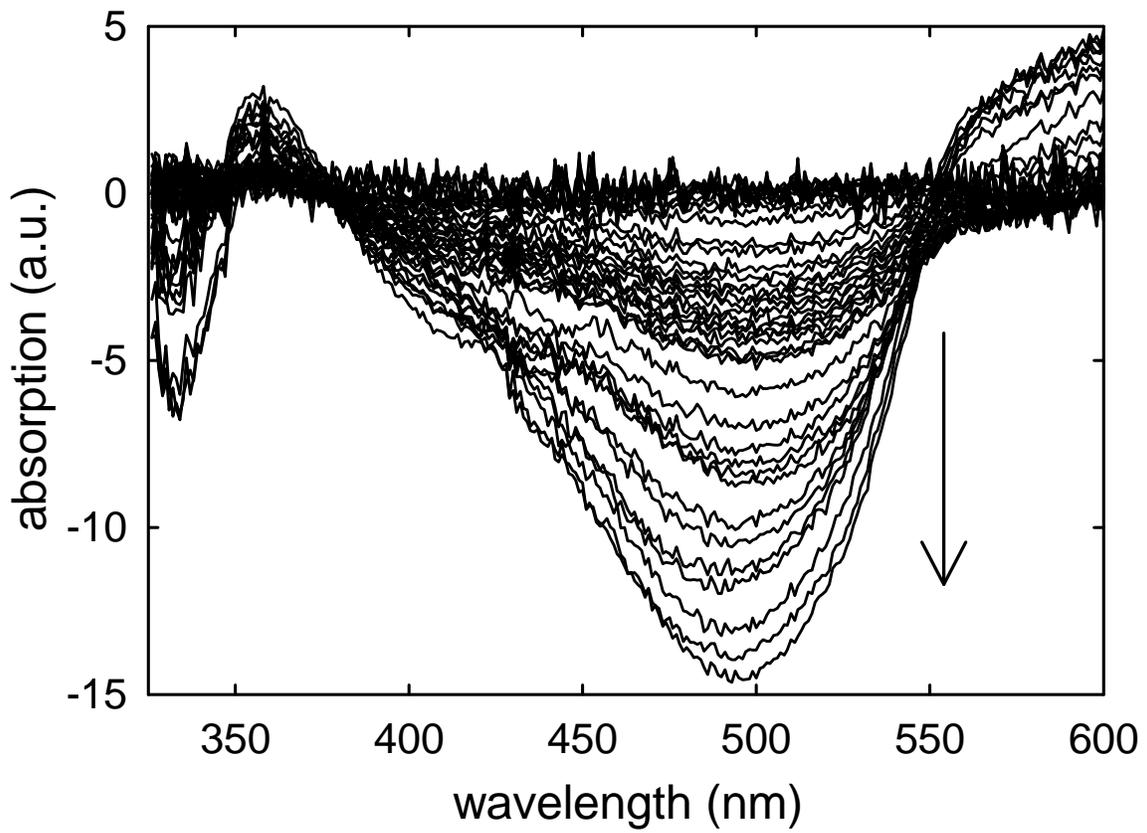

Fig 8

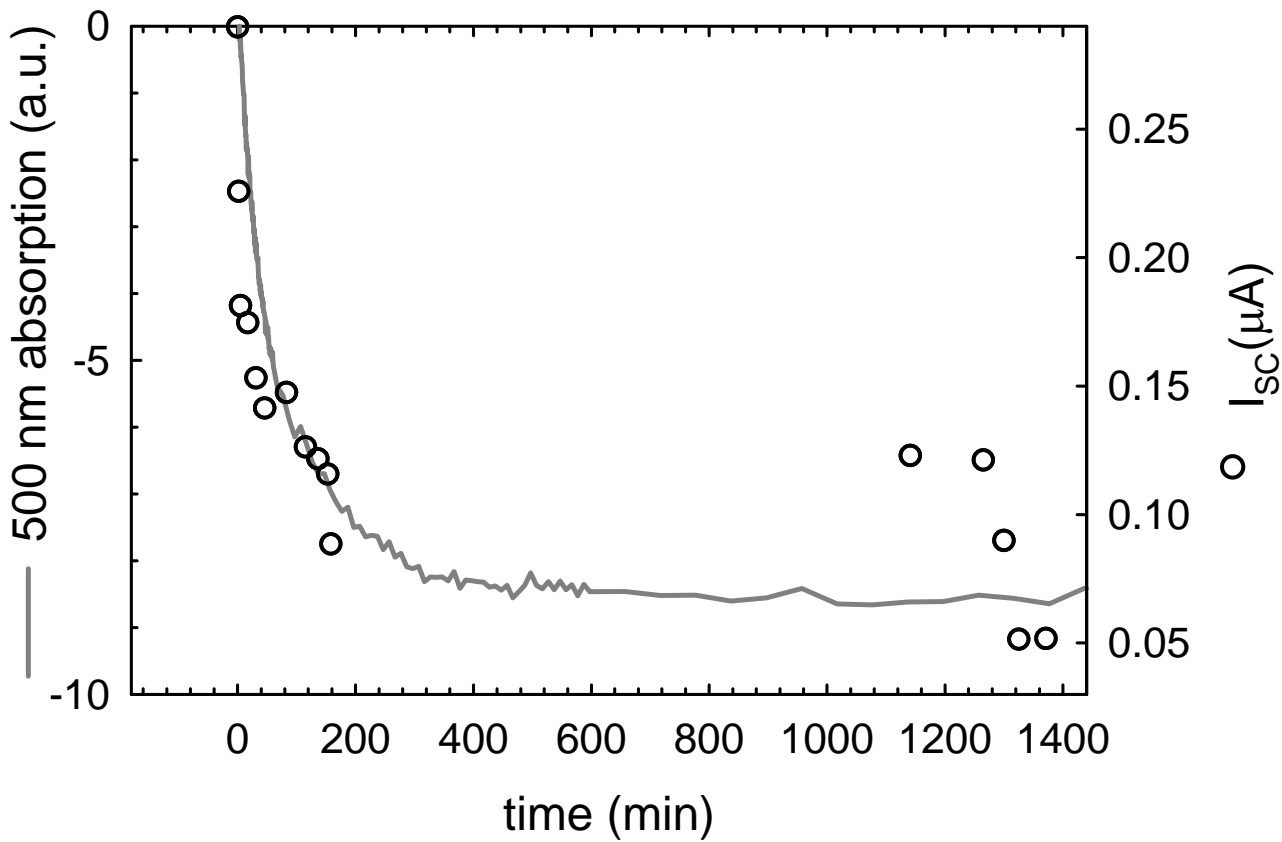

Fig 9